\documentclass[12pt]{article} 
\usepackage{graphicx}
\begin{document}
\baselineskip 10mm

\centerline{\bf Irradiation-induced suppression}
\centerline{\bf of the critical temperature in high-$T_c$ superconductors:}
\centerline{\bf Pair breaking versus phase fluctuations}

\vskip 2mm

\centerline{L. A. Openov}

\vskip 2mm

\centerline{\it Moscow Engineering Physics Institute
(State University)}
\centerline{\it 115409 Moscow, Russia}
\centerline{\it E-mail: opn@supercon.mephi.ru}

\vskip 4mm

\begin{quotation}

Experiments on the irradiation-induced suppression of the critical
temperature in high-$T_c$ superconductors are analyzed within the mean-field
Abrikosov-Gor'kov-like approach. It is shown that the experimental data
for YBa$_2$Cu$_3$O$_{7-\delta}$ single crystals can be quantitatively
explained by the pair breaking effects under the assumption of the combined
effect of potential and spin-flip scattering on the critical temperature and
with account for a non-pure $d$-wave superconducting order parameter.

\end{quotation}

\vskip 4mm


\vskip 6mm

Particle irradiation is a powerful tool that gives an opportunity to modify
the physical properties of superconductors.
Irradiation-induced defects act as effective pinning centers \cite{Civale},  thus causing
the critical current density to increase. Apart from the
practical benefits, irradiation effects may be used to probe the fundamental
characteristics of superconductors. For example, peculiarities of the
disorder-induced suppression of the critical temperature $T_c$ are expected
to depend on the pairing
mechanism and the symmetry of the superconducting order parameter
$\Delta({\bf p})$. In this respect, a study of the response of high-$T_c$
cuprates to the intentionally incorporated impurities or radiation defects
provides an indirect way to elucidate the cause of their unusual
normal and superconducting properties.
Among other things, depending on the symmetry of $\Delta({\bf p})$, clear
differences were predicted for the defect-induced variations of the
experimentally accessible characteristics such as $T_c$
\cite{Abrikosov,Openov1}, the density of states \cite{Pokrovsky},
the isotope coefficient \cite{Openov2}, the specific heat jump \cite{Openov3},
{\it etc}.

Various mechanisms of the disorder-induced $T_c$ suppression have been
considered, including, e. g., the pair breaking \cite{Radtke}, localization
\cite{Elesin}, and phase fluctuations \cite{EK} effects, {\it etc}.
The main problem here is that the disorder results not only in the decrease
of $T_c$ but also in the strong increase in the width of the superconducting
transition, $\Delta T_c$, so that the functional form of $T_c$ versus, e. g.,
the defect concentration $x_d$ appears to be poorly defined
at $T_c << T_{c0}$,
where $T_{c0}$ is the initial value of $T_c$ in the absence of the disorder.
In fact, the value of $\Delta T_c$ usually becomes comparable to the value of
$T_c$ at $T_c/T_{c0}\approx 0.3$ \cite{Elesin2,Tolpygo}.
While the measured $T_c$
versus $x_d$ curve in high-$T_c$ cuprates was commonly observed to be
approximately linear at $T_c/T_{c0} > 0.3$ \cite{Elesin2}, the details of
$T_c(x_d)$ behavior at $T_c/T_{c0} << 1$ remained unclear.

In a recent paper \cite{Rullier}, Rullier-Albenque {\it et al.} reported the
results of experimental studies of $T_c$ degradation under electron
irradiation of underdoped and optimally doped YBa$_2$Cu$_3$O$_{7-\delta}$
single crystals. They have measured $T_c$ and in-plane resistivity
$\rho_{ab}$ in a very broad range of $x_d$, the value of $x_d$ being
proportional to $\Delta \rho_{ab}$, the increase in $\rho_{ab}$ upon
irradiation. The authors of Ref. \cite{Rullier} succeeded in creation of an
extremely uniform distribution of radiation defects over the sample, so that
the value of $\Delta T_c$ never exceeded 5 K. Moreover, the value of
$\Delta T_c$ did not increase monotonously with radiation dose but had a
maximum at $T_c/T_{c0}\approx 0.3$ and next decreased again down to
$\Delta T_c < 1$ K at the highest dose for which the resistive
superconducting transition was still observed at $T_c\approx 1$ K. So, the
dependence of $T_c$ on $\Delta\rho_{ab}$ (or $x_d$) was obtained with an
excellent accuracy from $T_c/T_{c0}=1$ down to $T_c/T_{c0}=0$ (or, at least,
$T_c/T_{c0}\sim 10^{-2}$).

It was found in Ref. \cite{Rullier} that $T_c$ unexpectedly decreased
quasilinearly with $x_d$ in the {\it entire} range from $T_{c0}$ down to
$T_c=0$. Having
compared the results obtained with the predictions of Abrikosov-Gor'kov (AG)
pair breaking \cite{AG} and Emery-Kivelson phase fluctuations \cite{EK}
theories, the authors of Ref. \cite{Rullier} arrived at a conclusion that the
experimental data are at variance with AG theory and point to a significant
role of phase fluctuations of the order parameter in high-$T_c$
superconductors.

To compare the pair breaking theory with the experiment, the authors of
Ref. \cite{Rullier} made use of the AG formula \cite{AG} for a $d$-wave
superconductor (we set $\hbar=1$ hereafter)
\begin{equation}
\ln(T_{c0}/T_c)=\Psi(1/2+1/4\pi T_c\tau)-\Psi(1/2),
\label{Tc1}
\end{equation}
where $\Psi(z)$ is the digamma function and $\tau$ is the electron
scattering time \cite{Error},
$\tau^{-1}\propto x_d \propto \Delta \rho_{ab}$. This formula gives a
negative curvature of the $T_c$ versus $\Delta \rho_{ab}$ curve, contrary to
the experimental observations. Note, however, that, first, the
symmetry of $\Delta({\bf p})$ in YBa$_2$Cu$_3$O$_{7-\delta}$ is different
from pure $d$-wave due to an orthorombic lattice distortion \cite{Kouznetsov}
and, second, irradiation
may result in appearance of spin-flip scatterers along with potential ones
since radiation defects created in CuO$_2$ planes disturb antiferromagnetic
correlations between copper spins. The AG-like formula that accounts for both
those effects reads \cite{Openov1,Openov4}
\begin{eqnarray}
&&\ln\left(\frac{T_{c0}}{T_c}\right)=(1-\chi)\left[\Psi\left(\frac{1}{2}+%
\frac{1}{2\pi T_c\tau_s}\right)-\Psi\left(\frac{1}{2}\right)\right]
\nonumber \\
&&+\chi\left[\Psi\left(\frac{1}{2}+\frac{1}{4\pi T_c}\left(\frac{1}{\tau_p}+%
\frac{1}{\tau_s}\right)\right)-\Psi\left(\frac{1}{2}\right)\right]~,
\label{Tc2}
\end{eqnarray}
where $\tau_p$ and $\tau_s$ are scattering times due to potential and
spin-flip scatterers, respectively, the coefficient
\begin{equation}
\chi=%
1-\langle\Delta({\bf p})\rangle^2_{FS}/\langle\Delta^2({\bf p})\rangle_{FS}
\label{chi}
\end{equation}
is a measure of the degree of in-plane anisotropy of $\Delta({\bf p})$,
$\langle ... \rangle_{FS}$ means the Fermi surface (FS) average. The range
$0\leq\chi\leq 1$ covers the cases of isotropic $s$-wave
($\Delta({\bf p})$=const, $\chi=0$), $d$-wave
($\langle\Delta({\bf p})\rangle_{FS}=0$, $\chi=1$), and mixed $(d+s)$-wave
or anisotropic $s$-wave ($0<\chi<1$) symmetries of $\Delta({\bf p})$.

In fact, the assumption about the combined effect of potential and spin-flip
scatterers on $T_c$ and account for a non-pure $d$-wave $\Delta({\bf p})$ in
YBa$_2$Cu$_3$O$_{7-\delta}$ (i. e., $\chi \ne 1$) allow for a quantitative
explanation of the experimental data \cite{Rullier} within the modified pair
breaking AG-like theory \cite{Openov5}, without resorting to phase
fluctuations effects \cite{EK}. Fig. 1 shows the measured $T_c/T_{c0}$ versus
$\Delta \rho_{ab}$ taken from Ref. \cite{Rullier} along with theoretical
curves computed with Eq. (\ref{Tc2}) for $\chi=0.9$ and various values of
the coefficient
\begin{equation}
\alpha=\tau_s^{-1}/(\tau_p^{-1}+\tau_s^{-1})
\label{alpha}
\end{equation}
that specifies the relative contribution of spin-flip scatterers to the total
scattering rate. Here we represent the scattering time in
terms of the in-plane residual resistivity $\rho_0$ obtained by the
extrapolation of $\rho_{ab}(T)$ to $T=0$,
\begin{equation}
\tau_p^{-1}+\tau_s^{-1}=(\omega_{pl}^2/4\pi)\rho_0 ~ ,
\label{rho}
\end{equation}
where $\omega_{pl}$ is the plasma frequency, see Refs. \cite{Radtke} and
\cite{Openov4}. We also make use of the fact that $\rho_0=\Delta\rho_{ab}$
in a very good approximation \cite{Rullier}. From Fig. 1 one can see that
at $\chi=0.9$ and $\omega_{pl}=0.75$ eV the quasilinear
experimental dependence of $T_c$ on $\Delta\rho_{ab}$ in YBa$_2$Cu$_3$O$_7$
is quantitatively reproduced at $\alpha=0\div 0.01$.

We emphasize that the quantity $\omega_{pl}$ that enters Eq. (\ref{Tc2}) for
$T_c$ through the relation (\ref{rho}) should be considered as just a
characteristic energy which does not necessarily coincide with the value of
the plasma frequency determined by, e. g., the optical spectroscopy.
Based on general grounds, one could expect $\omega_{pl}\sim 1$ eV. In this
respect, although our choice of $\omega_{pl}=0.75$ eV is, to some extent,
arbitrary, the change in $\omega_{pl}$ results just in the change of the best
fitting values of $\chi$ and $\alpha$. For example, $\chi\approx 0.8$ and
0.6, $\alpha=0.04\pm 0.01$ and $0.045\pm 0.01$ for $\omega_{pl}=0.8$ and
1.0 eV, respectively, see Figs. 2 and 3. Meanwhile, for $\chi=1$, i. e., for
pure $d$-wave symmetry of $\Delta({\bf p})$, the experimental data cannot be
described at any value of $\omega_{pl}$, see Fig. 4. This is not surprising
because of the orthorombic crystal structure of YBa$_2$Cu$_3$O$_{7-\delta}$
which excludes the pure $d$-wave symmetry of $\Delta({\bf p})$ and points to
an admixture of the $s$-wave component to $d$-wave, so that $\Delta({\bf p})$
is of $(d+s)$-wave or $(d+is)$-wave type \cite{Kouznetsov}.
So, the experimental data \cite{Rullier}
for YBa$_2$Cu$_3$O$_7$ single crystals can be quantitatively
explained by the pair breaking theory taking a non-pure $d$-wave
$\Delta({\bf p})$ and the combined effect of potential and spin-flip
scatterere on $T_c$ into account.

As for the underdoped single crystals YBa$_2$Cu$_3$O$_{6.6}$, the
experimental dependence \cite{Rullier} of $T_c/T_{c0}$ versus
$\Delta\rho_{ab}$ is close to that for YBa$_2$Cu$_3$O$_7$ and can be fitted
within the same approach at similar values of $\omega_{pl}$, $\chi$, and
$\alpha$. The discussion of the probable effect of the oxygen content, i. e.,
the hole concentration, on the value of $\omega_{pl}$, the gap anisotropy,
and the relative amount of spin-flip scatterers in the sample is, however,
beyond the scope of this paper.

Note that $\chi < 1$ not only for a mixed $(d+s)$-wave $\Delta({\bf p})$,
but also for an anisotropic $s$-wave $\Delta({\bf p})$.
Recently the $d$-wave symmetry of
$\Delta({\bf p})$ in hole-doped cuprate superconductors
\cite{Tsuei} has been doubted by several authors
(see, e. g., Refs. \cite{Brandow,Zhao}).
The re-analysis of the results obtained by
the angle-resolved photoemission spectroscopy, the Fourier transform scanning
tunneling spectroscopy, the low-temperature thermal conductivity, {\it etc.},
including the phase-sensitive techniques, has shown that the combined data
agree quantitatively with the
extended $s$-wave symmetry \cite{Brandow,Zhao}. Making use of the fit
\cite{Zhao} $\Delta(\theta)=24.5(\cos4\theta+0.225)$ meV to single-particle
tunneling spectra of YBa$_2$Cu$_3$O$_{7-\delta}$, the angle $\theta$ being
measured from the Cu-O bonding direction, we have $\chi\approx 0.9$ for
YBa$_2$Cu$_3$O$_{7-\delta}$. It follows from the fits presented in
Ref. \cite{Zhao} that even more lower value of $\chi$ may be
expected for Bi$_2$Sr$_2$CaCu$_2$O$_{8+y}$. In this respect, it would be very
interesting to study the behavior of $T_c$ versus $\rho_{ab}$ in this and
other high-$T_c$ cuprates down to $T_c=0$.

Finally, a note is in order about one more argument presented in
Ref. \cite{Rullier} in favour of the phase fluctuations theory and against
the pair-breaking mechanism of $T_c$ suppression in high-$T_c$ cuprates.
According to Ref. \cite{Rullier}, the positive curvature of the
$T_c(\Delta\rho_{ab})$ curve is necessarily required to explain the maximum
of the transition width $\Delta T_c$ as a function of $\Delta\rho_{ab}$ that
was experimentally observed at $T_c/T_{c0}\approx 0.3$. Note, however, that,
first, this argument is incompatible with the experimental data
themselves since the curvature of the {\it measured} $T_c(\Delta\rho_{ab})$
dependence is (with a few exceptions) close to zero in the whole range of
$\Delta\rho_{ab}$ and, respectively, in the whole range of $T_c/T_{c0}$,
including the region near
$T_c/T_{c0}\approx 0.3$. Second, the line of reasoning in Ref. \cite{Rullier}
is based on a naive assumption that
$\Delta T_c(x_d) \propto x_d(dT_c/dx_d)$. Such an assumption is at
least questionable for the resistive superconducting transition whose
critical temperature and width are determined by the zero-resistance path and
the uniformity of the defect distribution, respectively. Besides, the value
of $\Delta T_c$ depends on a specific criterion used for its evaluation
from the curve $\rho_{ab}(T)$. Thus, the knowledge of the function
$T_c(x_d)$ alone is obviously insufficient to draw the definite
conclusions about the function $\Delta T_c(x_d)$, and vice versa.

We note that the phase fluctuations theory \cite{EK} goes beyond
the standard mean-field theory and implies that the so called
pseudogap \cite{Timusk} is a precursor to superconductivity. This
contradicts the experiments which give evidence for interplay
between competing and coexisting (superconducting and
non-superconducting) ground states, see, e. g., Ref. \cite{Alff}.
We note also that the AG-like pair breaking approach is based on
the BCS-Bogolubov mean-field theory that seems to describe the
spatial-momentum quasiparticle states in high-$T_c$ cuprates, at
least in the optimally doped samples such as, e. g.,
YBa$_2$Cu$_3$O$_7$, rather well \cite{McElroy,Matsui}.

In summary, we have shown that experiments on the irradiation-induced $T_c$
suppression in YBa$_2$Cu$_3$O$_{7-\delta}$ can be quantitatively explained
within the AG-like pair breaking mean-field theory under the assumption of
the combined effect of potential and spin-flip scattering on $T_c$ and
with account for a nonzero Fermi surface average of the superconducting order
parameter, without resorting to phase fluctuations effects. One can
not exclude, however, a possibility that the latter become important at
$T_c\rightarrow 0$, i. e., in the very vicinity of the
superconductor-insulator transition.

\vskip 2mm

I am grateful to A. V. Kuznetsov for assistance.

\newpage

\includegraphics[width=\hsize]{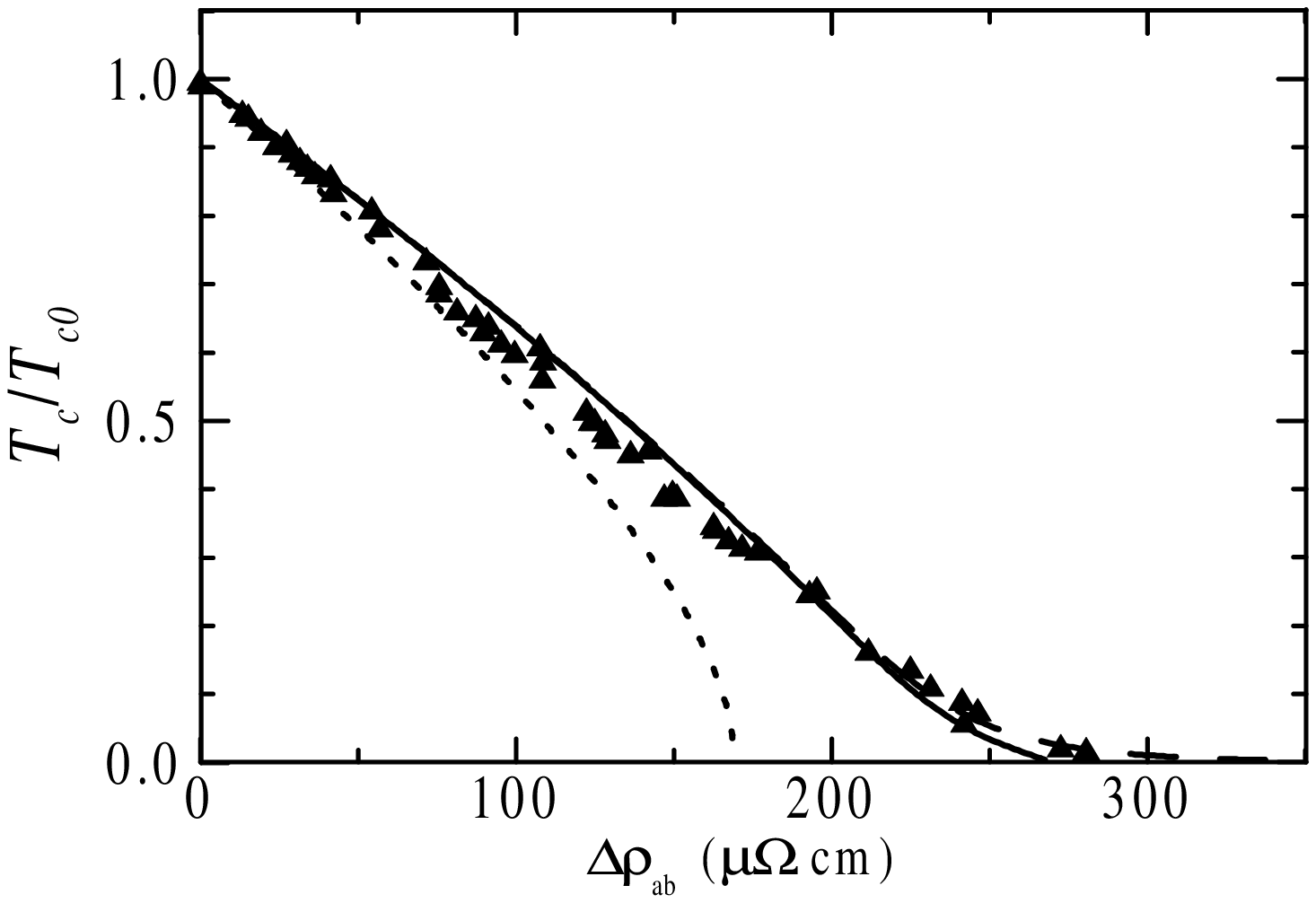}

\vskip 6mm

Fig. 1. $T_c/T_{c0}$ versus $\Delta\rho_{ab}$ in electron irradiated
YBa$_2$Cu$_3$O$_7$ crystals. Experiment \cite{Rullier} (triangles).
Theory, Eqs. (\ref{Tc2}) - (\ref{rho}), for $\omega_{pl}=0.75$ eV,
$\chi=0.9$, and $\alpha=0$ (dashed line), 0.01 (solid line), and 1
(dotted line).

\newpage

\includegraphics[width=\hsize]{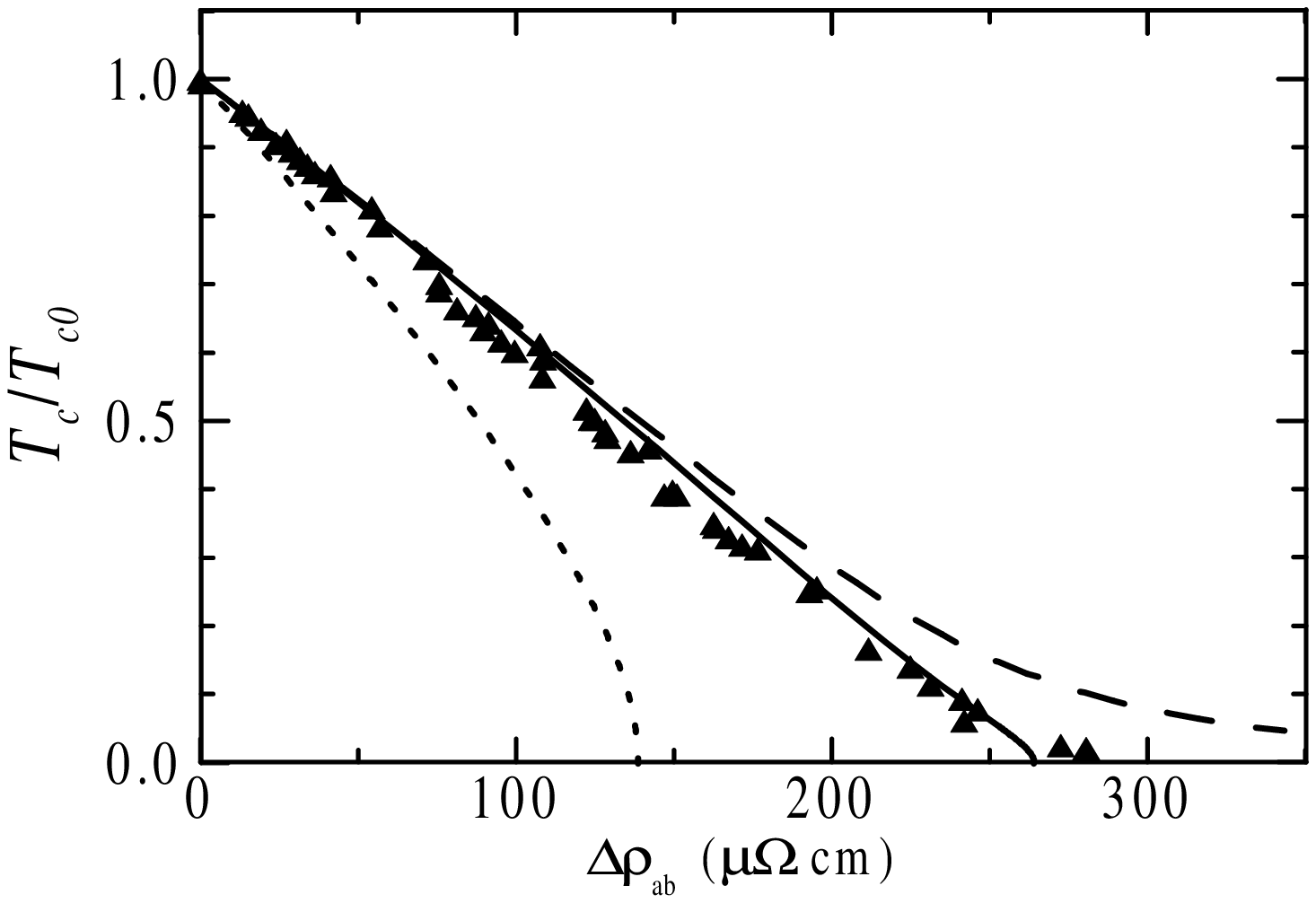}

\vskip 6mm

Fig. 2. The same as in Fig. 1 for $\omega_{pl}=0.8$ eV, $\chi=0.8$, and
$\alpha=0$ (dashed line), 0.04 (solid line), and 1 (dotted line).

\newpage

\includegraphics[width=\hsize]{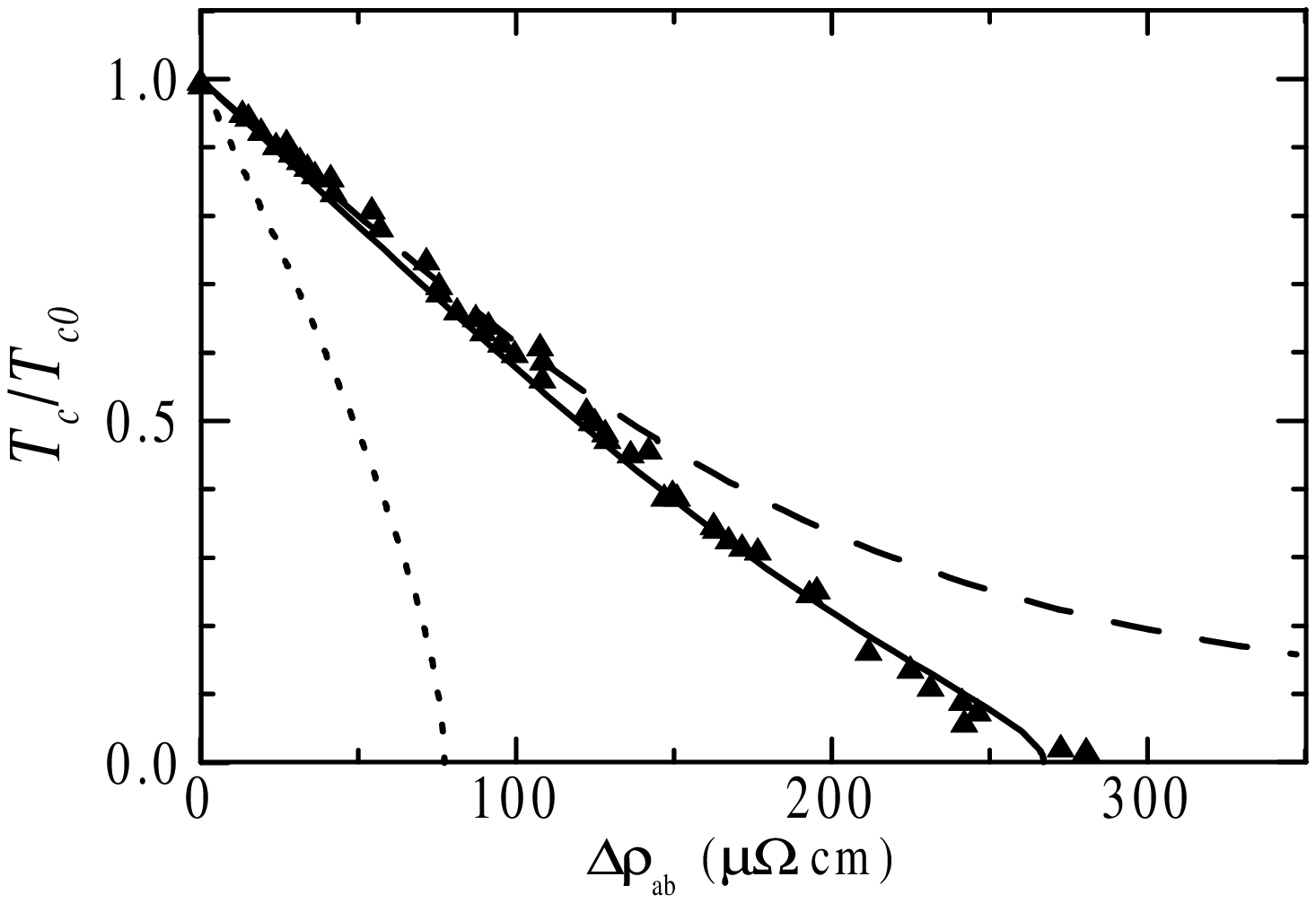}

\vskip 6mm

Fig. 3. The same as in Fig. 1 for $\omega_{pl}=1.0$ eV, $\chi=0.6$, and
$\alpha=0$ (dashed line), 0.045 (solid line), and 1 (dotted line).

\newpage

\includegraphics[width=\hsize]{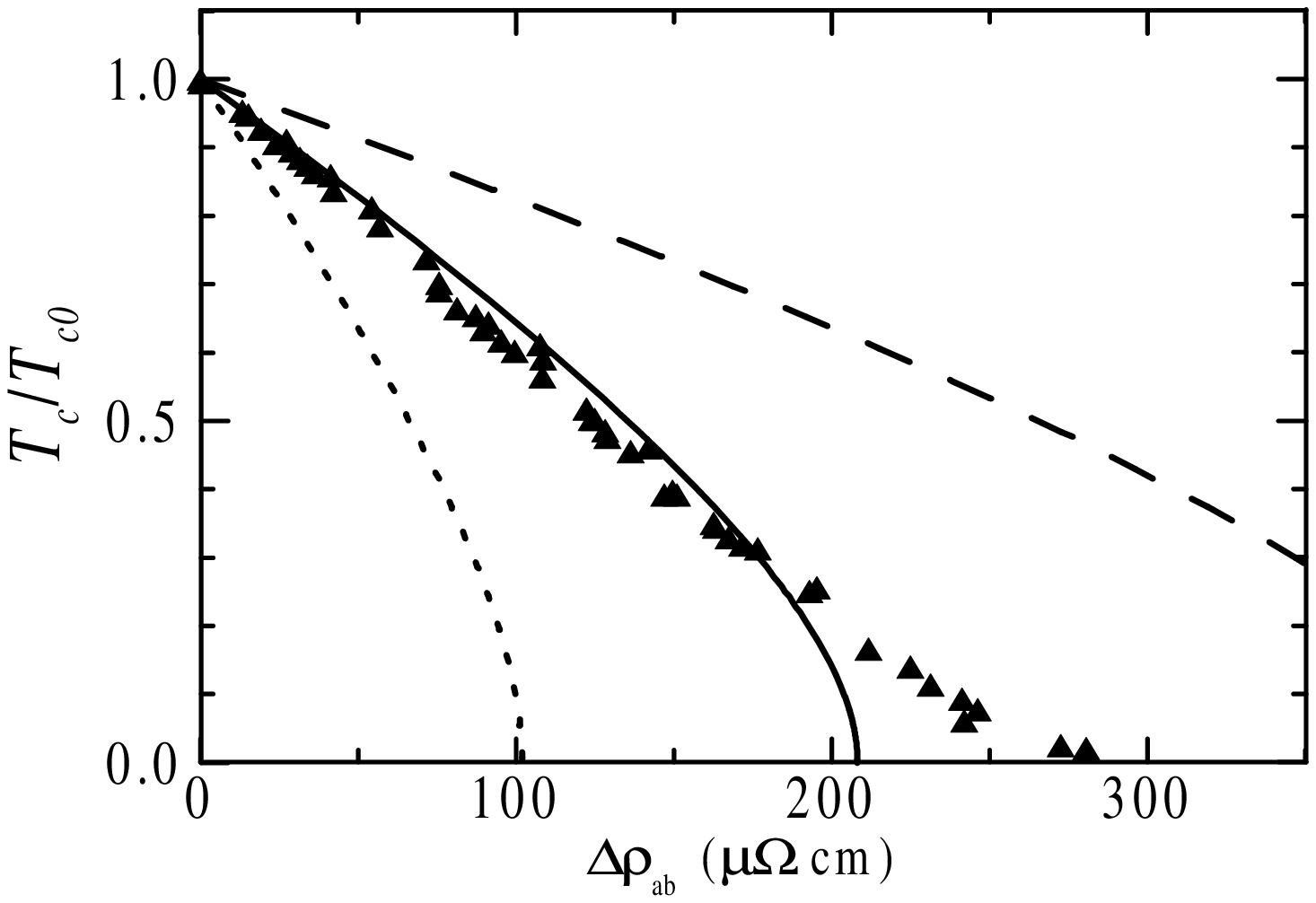}

\vskip 6mm

Fig. 4. The same as in Fig. 1 for $\chi=1$ and $\omega_{pl}=0.5$ eV
(dashed line), 0.7 eV (solid line), and 1 eV (dotted line),
see Ref. \cite{Note}.


\begin{thebibliography}{5}
\bibitem{Note} Note that in a $d$-wave superconductor ($\chi$=1), the value
of $T_c$ does not depend on $\alpha$, see Eqs. (\ref{Tc2}) and (\ref{alpha}).
\end{thebibliography}
\end{document}